\documentclass[10pt]{iopart}
\usepackage{graphicx}
\usepackage{amsfonts, amssymb}

\begin{document}

\title{Waves and rays in plano-concave laser cavities, Part~II: a semiclassical approach.}

\author{A. Pascal$^{1}$, S. Bittner$^{2}$, B. Dietz$^{3}$, A. Trabattoni$^{1}$, C. Ulysse$^{4}$, M. Romanelli$^{5}$, M. Brunel$^{5}$,  J. Zyss$^{1}$, and M. Lebental,$^{1}$}

\address{$^{1}$ Laboratoire de Photonique Quantique et Mol\'eculaire, CNRS UMR 8537, Institut d'Alembert FR 3242, Ecole Normale Sup\'erieure de Cachan, Universit\'e Paris-Saclay, F-94235 Cachan, France.}
\address{$^{2}$ Department of Applied Physics, Yale University, New Haven, Connecticut 06520, USA.}
\address{$^{3}$ School of Physical Science and Technology, and Key Laboratory for Magnetism and Magnetic Materials of MOE, Lanzhou University, Lanzhou, Gansu 730000, China.}
\address{$^{4}$ Centre de Nanosciences et de Nanotechnologies, Universit\'e Paris-Saclay, F-91460 Marcoussis, France.}
\address{$^{5}$ Institut de Physique de Rennes, Universit\'e Rennes I - CNRS UMR 6251,
Campus de Beaulieu, 35042 Rennes Cedex, France.}

\ead{melanie.lebental@ens-cachan.fr}


\begin{abstract}
This second paper on the Fabry-Perot cavity presents a semi-classical approach, which means that we consider the transition from wave optics to geometrical optics. The basic concepts are the periodic orbits and their stability. For the plano-concave Fabry-Perot cavity in the paraxial approximation, the derivation of the trace formula demonstrates that the spectrum is based only on the axial periodic orbit and its repetitions. Experiments with micro-lasers illustrate the relation to periodic orbits. The methods presented in this paper are not limited to laser cavities and can be applied to a large range of wave systems.
\end{abstract}

\pacs{} 
\submitto{\EJP}

\section{Introduction}

In this second contribution on the Fabry-Perot cavity, we propose to develop a complementary approach, called \emph{semi-classical physics}, which turns out to be of broad validity and applicability \cite{brack,stoeckmann-livre}. These techniques were developed in the early birth of quantum physics \cite{stone-physicstoday} and were extended to almost all domains where a transition from a wave description to geometrical concepts makes sense, from nuclear physics \cite{bohigas} to acoustics \cite{acoustique-mortessagne} and hydrodynamics \cite{chaos-hydro}, and even in quantum gravity \cite{gravitequantique-bianchi}. In electromagnetism, it simply corresponds to the transition from wave optics to geometrical optics.\\

In this article, we illustrate these methods on the well-known Fabry-Perot cavity and describe its features from an unusual perspective. The theory is compared with experiments based on microcavity lasers \cite{matsko-livre,Cao2015}. These microlasers are not to be viewed simply as downscaled versions of large laser cavities, such as studied in Part I \cite{PartI}, but are also permit a wealth of quasi-unlimited cavity shapes, which are defined by their external boundary. Such shapes can be tailored to any geometry of interest, either in a planar two-dimensional geometry as studied in this article or in three dimensions, the latter an emerging field beyond the scope of this article \cite{OE-3D}. Optical microcavities have been studied in great depth over the last two decades by means of a dual perspective based on ray optics and wave optics, in the same spirit as for the macroscopic cavities discussed in Part I. This dual approach has led to a confrontation of the two points of view, as in Part I, driven by the need to ensure compatibility and to explore the relevant connections between these two approaches. The main differences to Part I are, first, that the relations between the spectrum of the cavity and periodic orbits are explored and, second, that modes beyond a paraxial description are observed experimentally.\\

The toolbox for connecting ray optics with a modal picture has been developed since the seventies within the framework of semiclassical physics \cite{brack}. Initially triggered by nuclear physics where statistical regularities were shown to underpin the complexity of the spectra of nuclei \cite{bohigas}, it was later applied to connect the spectra of closed microwave cavities with the periodic orbits of classical billiards \cite{graf1992}, and more recently to the case of open optical cavities where radiation is only partially confined by reflections at the dielectric interfaces \cite{PRE-trace}.
The density of states, a familiar notion in classical as well as quantum physics, in optics and solid state physics alike, can be connected to the set of all periodic orbits, which are purely classical quantities, by virtue of the celebrated \emph{trace formula}, which has been proposed in the seventies \cite{balian,gutzwiller}, and progressively extended to a broad variety of quantum- and wave-mechanical systems.\\

This trace formula is the cornerstone of a semiclassical approach to resonators, but its implementation is far from obvious. Indeed the beauty and simplicity of its formulation must not hide the technical difficulties that are met in practice, even in the simple cases on which we will concentrate hereafter. Its ingredients are on the one hand the solutions of Maxwell's equations to obtain the density of electromagnetic modes, and on the other hand geometric considerations to obtain the periodic orbits, which are usually non-trivial, depending on the nature of the resonator shape. A vital role is played by the stability of the periodic orbits, which determines the expansion amplitudes and thus their weight of contribution to the trace formula. In order to put our approach on simple but solid grounds, we start with the illustrative case of the plane-parallel Fabry-Perot cavity, where the identification of the periodic orbits is straightforward, allowing to concentrate on the methodology and its physical significance. We then move on to plano-concave Fabry-Perot microcavities like those considered in Part I.\\

One main result, which connects in an essential manner the two parts of this series, is the recognition of the role of modal degeneracy, here based on the full calculation of the stability-related amplitudes in the trace formula. The condition for such degeneracy is shown to take the same form as already established in Part I [see Eq.~(39)], but is derived from a different perspective and with different tools. This connection is easily generalized to other systems and thus of broad relevance. In that light, the dips in Fig.~10 of Part I are interpreted as a local change of stability of the repetitions of the axial periodic orbit. \\

The outline is the following. The experiments are sketched in Sec.~\ref{sec:experiences-intro}, so as to introduce the main issues which are discussed in this paper. The theoretical background is then introduced in Sec.~\ref{sec:theorie} with the simplest example of the plane-parallel Fabry-Perot cavity, then applied to the specific case of plano-concave cavities. The experimental results are finally presented and compared to the theory in Sec.~\ref{sec:experiences}.

\section{Experiments and main issues}\label{sec:experiences-intro}

\begin{figure}[tb]
\begin{center}
\includegraphics[height = 4 cm]{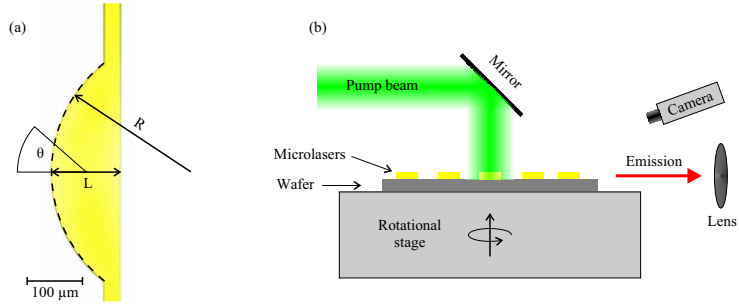}
\end{center}
\caption{(a) A photograph with an optical microscope (top view) in real colors of a plano-concave microlaser. The width of the cavity is $L = 125~\mu$m and the radius of the concave boundary $R = 250~\mu$m in this example ($\rho = 1/2$). The far-field observation angle is $\theta$. (b) Scheme of the optical characterization setup (side view, not to scale).}
\label{fig:geomAndSetup}
\end{figure}

The plano-concave Fabry-Perot cavities consist of a plane mirror within a distance $L$ from a convex mirror of radius $R$. A microscope image of such a cavity and the notations are presented in Fig.~\ref{fig:geomAndSetup}(a). In this paper, as in Part I, we focus on the case $L / R \equiv \rho < 1$, corresponding to a stable periodic orbit along the principal axis. The question of stability will be discussed in Sec.~\ref{sec:stabilite}.

\subsection{Microlaser experiments}

The microlasers are made of a single polymer-doped layer, which is spin-coated on a silica-on-silicon wafer. The passive polymer host is poly(methyl methacrylate) (PMMA) which is transparent for visible light. It is doped at a concentration of \mbox{$5$ weight \%} with the commercial laser dye DCM \footnote{[2-[2-[4-(dimethylamino)phenyl]ethenyl]-6-methyl-4H- pyran-4-ylidene]-propanedinitrile, DCM, provided by Exciton.}. The $2~\mu$m silica buffer layer has a lower refractive index than the $650$~nm thick polymer-doped layer. It thus ensures that the laser light propagates in two dimensions within the plane of the polymer-doped layer \cite{PRA-simples}. The cavities are then patterned by electron beam lithography, which provides a nanoscale etching quality.\\

The microlasers are pumped individually with a frequency doubled Nd:YAG laser ($532$~nm, $500$~ps, $10$~Hz). The pump beam is incident from the top as shown in Fig.~\ref{fig:geomAndSetup}(b) and uniform at the cavity scale. The emission is collected in the substrate plane and transmitted to a spectrometer coupled to a cooled CCD camera with a spectral resolution of about $0.03$~nm. For example, considering the plane-parallel Fabry-Perot laser of Fig. 2(a), one can measure the typical experimental spectrum of Fig. 2(b). It features a single comb of quasi-equidistant peaks. \\

\begin{figure}[tb]
\begin{center}
\includegraphics[width = 1\linewidth]{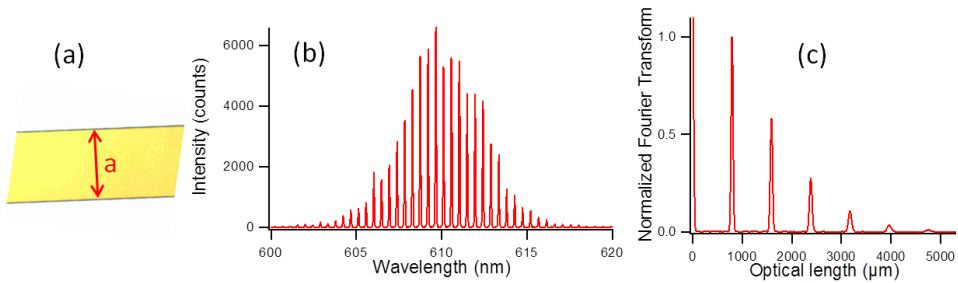}
\end{center}
\caption{Plane-parallel Fabry-Perot micro-laser. (a) Photograph with an optical microscope in real colors. The only periodic orbit is indicated in red. (b) Experimental spectrum of a plane-parallel Fabry-Perot cavity with $L=250~\mu$m. (c) Fourier transform of the spectrum in (b).}
\label{fig:plan-plan}
\end{figure}

These are not table-top experiments, since a (short) pulsed laser is required for pumping solid-state dye lasers, and the fabrication of the devices involves electron-beam lithography to ensure the verticality of the cavity walls. However these facilities are conveniently affordable by an average laboratory.

\subsection{Main issues}

In some respects, a laser can exhibit intriguing and counter-intuitive features. However we will see that most of its properties can be simply explained by studying fundamental issues of the passive cavity, without gain. \\

Which modes are lasing ? Why these modes and not others, since there are so many modes in the passive cavity ?\\
For cavities much larger than the wavelength, the wave properties are based on the periodic orbits of the passive cavity, which are closed periodic geometrical trajectories \cite{PRE-trace}. If the gain is enhanced on a specific periodic orbit, the lasing modes tend to localize on this orbit \cite{doya}.\\

What does it mean that a cavity is called \emph{stable} or \emph{unstable} ?\\
The stability property does not refer to the whole cavity, but to a single periodic orbit. In general, the Fabry-Perot cavities are used in paraxial conditions and the gain is localized along the principal axis. Hence, we must consider the axial periodic orbit, which is drawn with a continuous red line in Fig.~\ref{fig:differentes-orbites}a. It is stable if $0<L/R<1$, which means that the potential is confining and that rays tend to remain in its vicinity. A precise definition of stability is given in Sec.~\ref{sec:stabilite}.\\

In Fig. 10 of Part I, why do the lasing modes present such a different behavior for specific ratios $L/R$ ?\\
As explained in Part I, these specific ratios correspond to a degeneracy of modes in the paraxial approximation. From a ray-dynamical point of view, the potential is not perfectly confining at these specific ratios $L/R$ and some repetitions of the axial periodic orbit are only marginally stable, which means that the light can explore a larger area around the axial orbit. \\

The aim of this paper is to give precise answers to each of these questions.

\section{Wave vs classical mechanics: the trace formula} \label{sec:theorie}

The trace formula was developed in the seventies \cite{balian,gutzwiller} and stands out due to its ability to express a spectrum, which is a notion that intrinsically belongs to wave mechanics alone, in terms of geometric entities which are pertaining to classical physics. It is hence an important connection between wave physics and the geometrical world.\\

Sec.~\ref{sec:trace-generique} gives an overview of the trace formula, while the simplest case of the plane-parallel Fabry-Perot cavity is derived in Sec.~\ref{sec:trace-FP}. Since the stability of periodic orbits is of fundamental importance for the trace formula, the stability of orbits is discussed in detail in  Sec.~\ref{sec:stabilite}. Finally Sec.~\ref{sec:orbites-plano-concave} and Sec.~\ref{sec:trace-plan-concave} deal with the plano-concave cavity and discuss its periodic orbits and the trace formula for the axial orbit.

\subsection{Trace formula: Generic case} \label{sec:trace-generique}

\begin{figure}[tb]
\begin{center}
\includegraphics[height = 3.5 cm]{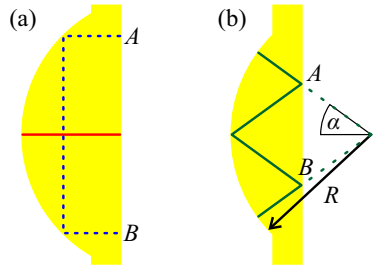}
\end{center}
\caption{Several examples of periodic orbits in a plano-concave cavity. (a) Cavity with $\rho = 0.7$. The continuous red line indicates the axial orbit and the dotted blue line the $C$-orbit. Both orbits are isolated and exist for a wide range of $\rho$. (b) Cavity with $\rho = 0.5$. One example of the family of the $W$-orbit is indicated. This family of marginally stable orbits only exists for $\rho = 0.5$.}
\label{fig:differentes-orbites}
\end{figure}

The density of states (i.e., the spectrum) is defined as
\begin{equation} d(k) = \sum_j \delta(k - k_j) \end{equation}
where the $\{k_j\}$ are the eigen-wavenumbers of the system. In the \emph{semiclassical limit} or geometrical-optics limit -- which means $ka \gg 1$ for $a$ a typical length of the system -- $d(k)$ tends to a sum over the periodic orbits ($po$) of that same system,
\begin{equation} \label{eq:trace-definition}
d(k) \propto \sum_{po} C_{po} \cos(k\mathcal{L}_{po} + \phi_{po}) \, .
\end{equation}
A periodic orbit is a classical, closed, and periodic trajectory. Figure \ref{fig:differentes-orbites} shows different periodic orbits in a plano-concave cavity. The summation in Eq.~(\ref{eq:trace-definition}) is performed over all the periodic orbits, with the amplitude $C_{po}$ depending only on geometric parameters of the orbit such as its stability and its length $\mathcal{L}_{po}$. Likewise, $\phi_{po}$ is a phase term which only relates to classical quantities as well. This expression has been established for integrable \cite{balian} and chaotic \cite{gutzwiller} cases in the semiclassical limit. The microlasers considered here have $kL \sim 1000$ and are thus well within the semiclassical regime where the trace formula applies. A pedagogical derivation of the trace formula is proposed in \cite{bogomolny-2003} for integrable systems (\S 3) and chaotic systems (\S 4). 

\subsection{Trace formula for the plane-parallel Fabry-Perot cavity} \label{sec:trace-FP}

Let us consider the simplest example of a plane-parallel Fabry-Perot cavity. There exists a single periodic orbit, see Fig.~\ref{fig:plan-plan}(a), and all its repetitions. It should be noted that this orbit is not isolated but actually belongs to a continuous family of parallel orbits, all of which have the same properties, but in the following we will only refer to it as single periodic orbit for simplicity's sake. The lengths of the primitive orbit and its repetitions are $\mathcal{L}_Q = Q \, 2 L$, with $Q \in \mathbb{N}^{\star}$ being the number of repetitions. Their corresponding amplitudes $C_Q$ are  smooth functions of $Q$, see Sec.~\ref{sec:stabilite}, and their phases $\phi_Q = 0\,[2\pi]$. Therefore the trace formula (\ref{eq:trace-definition}) reduces to
\begin{equation} \label{eq:trace-plan-plan}
d(k) \propto \sum_{Q\in\mathbb{N}^{\star}} \cos(2QL\,k) \propto \sum_{Q\in\mathbb{Z}^{\star}} e^{2 i Q \, L \, k}
\end{equation}
which is equivalent to a Dirac comb. This is also observed experimentally: in Fig.~\ref{fig:plan-plan}(b), the peaks are indeed equidistant, and the comb is modulated at a larger scale by the gain curve of the laser dye. The Fourier transform of expression (\ref{eq:trace-plan-plan}) is also a Dirac comb, which is peaked at every multiple of $\mathcal{L}=2L$, as shown in Fig.~\ref{fig:plan-plan}(c)\footnote{In practice, the Fourier transform is calculated from the spectrum expressed as a function of the wavenumber, and not of the wavelength, in consistency with formula (\ref{eq:trace-plan-plan}).}. For this experiment, $L = 250~\mu$m, so the first peak in the Fourier transform is expected at $2 L = 500~\mu$m, while it is observed at $(795 \pm 20)~\mu$m. In fact, the refractive index must be taken into account as the light travels through a dielectric medium. The trace formula is modified for dielectric resonators in a natural manner \cite{PRE-trace},
\begin{equation} \label{eq:trace-dielectrique}
d(k) \propto \sum_{po}r_{po} \, C_{po} \cos(n k \mathcal{L}_{po} + \phi_{po}) \, .
\end{equation}
The weighting factor is supplemented by $r_{po}$, the product of the Fresnel reflection coefficients corresponding to the reflections of the periodic orbit. It has a modulus of $|r_{po}| = 1$ for a periodic orbit which is confined by total internal reflection. For the orbit in the plane-parallel Fabry-Perot cavity, $r_{po} = (n - 1)^2 / (n + 1)^2 = 0.04$ due to two reflections at perpendicular incidence with $n = 1.5$ being the refractive index of the polymer cavity. The geometrical length of the periodic orbit is replaced by its optical length $n \mathcal{L} = 1.5 \cdot 500 = 750~\mu$m. This value is actually slightly lower than what is observed experimentally since the dispersion of the refractive index also needs to be taken into account \cite{PRA-simples}. Including all corrections, we use $n = 1.63$ for the data presented in this article.\\

A more usual -- and equivalent -- way to calculate the spectrum in Fig.~\ref{fig:plan-plan}(b) is the phase loop condition after a round trip of the light along the periodic orbit,
\begin{equation} \label{eq:retour-en-phase}
e^{i k \, n \mathcal{L}} = 1 \hspace{0.5cm} \Rightarrow \hspace{0.5cm} k_q = \frac{2 \pi}{n \mathcal{L}} \, q \hspace{0.5cm} q \in \mathbb{N}^{\star} \, .
\end{equation}

\subsection{Stability of periodic orbits}\label{sec:stabilite}

\begin{figure}[tb]
\begin{center}
\includegraphics[height = 4 cm]{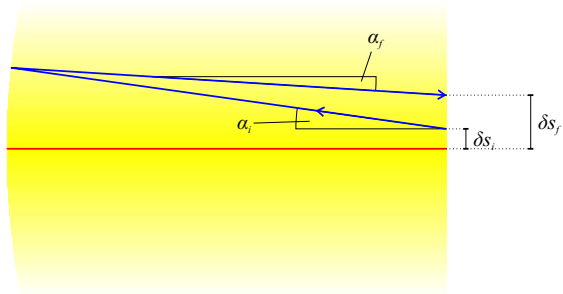}
\end{center}
\caption{Effect of a small perturbation on the axial periodic orbit (red). The blue trajectory starts from a position slightly displaced transversely by $\delta s_i$  and in a direction deviating by $\delta \alpha_i$. It arrives with a displacement $\delta s_f$ and a deviation $\delta \alpha_f$ after one round trip.}
\label{fig:monodromy}
\end{figure}

In the trace formula (\ref{eq:trace-definition}), the factor $C_{po}$ controls the weight of different periodic orbits in the spectrum. Its functional form depends on the stability of the considered periodic orbit, i.e., how the trajectory reacts to a small perturbation of the initial position or propagation direction. This stability can be calculated with the \emph{monodromy matrix}, which is equivalent to the ABCD matrix used in geometrical optics and in the framework of the paraxial approximation.\\

The theory of the monodromy matrix is described in \cite{brack,bogomolny-2003}. Here we focus on the calculation of the stability of the periodic orbits. As shown in Fig.~\ref{fig:monodromy}, the monodromy matrix $\mathcal{M}$ describes the effect of a small perturbation of the initial conditions, $(\delta s_i, \delta \alpha_i)$, after one round trip along the orbit at the linear order,
\begin{equation}
\left( \begin{array}{c} \delta s_f \\ \delta \alpha_f \end{array} \right) = \mathcal{M} \left( \begin{array}{c} \delta s_i \\ \delta \alpha_i \end{array} \right) \, .
\end{equation}
In the case of two-dimensional cavities considered here, the monodromy matrix $\mathcal{M}$ is a $2 \times 2$ matrix of determinant $1$. It can be calculated as the product of elementary matrices
\begin{equation}
\Pi(L) = \left( \begin{array}{cc} 1 & L\\
0 & 1\end{array} \right)
\end{equation}
for the propagation in a straight line of length $L$ and
\begin{equation}
\mathcal{R}(R, \chi) = \left( \begin{array}{cc} -1 & 0 \\
\frac{2}{R \cos(\chi)} & -1 \end{array} \right)
\end{equation}
for the reflection with an angle of incidence $\chi$ with respect to the surface normal on a surface with radius of curvature $R$. For instance, for the isolated periodic orbit along the axis of a concave-concave Fabry-Perot cavity of length $L$, the monodromy matrix is $\mathcal{M} = \Pi(L) \, \mathcal{R}(R_1,0) \, \Pi(L) \, \mathcal{R}(R_2,0)$, where $R_1$ and $R_2$ are the radii of curvature of the mirrors. 

As the monodromy matrix $\mathcal{M}$ is a 2$\times$2 matrix of determinant $1$, there are three possible cases for its two eigenvalues $\Lambda_+$ and $\Lambda_-$.
\begin{itemize}
\item $\Lambda_+$ and $\Lambda_-$ are real. Since the determinant is $1$, their product $\Lambda_+ \Lambda_- = 1$ and so $\Lambda_+ = 1/\Lambda_- > 1$. The Lyapunov exponent $\lambda$ of the orbit is defined via $\Lambda_+ = e^{\lambda}$. The larger $\lambda$, the more \emph{unstable} is the periodic orbit. In this case, the amplitude is
\begin{equation}\label{eq:cp-instable}
    C_{po} \propto \frac{1}{\sinh(\lambda / 2)} \, .
\end{equation}
\item $\Lambda_+=\Lambda_-=1$. The periodic orbit is \emph{marginally stable}. This is for instance the case of the periodic orbit of the plane-parallel Fabry-Perot cavity.
\item The eigenvalues are complex conjugated with a modulus $1$: $\Lambda_+ = 1 / \Lambda_- = e^{i \Phi}$. In this case, the periodic orbit is \emph{stable} and the amplitude is
\begin{equation}\label{eq:cp-stable}
    C_{po} \propto \frac{1}{\sin(Q \Phi / 2)}
\end{equation}
where $Q \in \mathbb{N}^{\star}$ is the number of repetitions of the periodic orbit.
\end{itemize}
Actually, there are two other cases for $\Lambda_+$ and $\Lambda_-$ both having negative signs. Then their amplitudes $C_{po}$ are slightly different (see Ref.~\cite{brack} Table 5.1).\\

Because the monodromy matrix $\mathcal{M}$ is a $ 2 \times 2$ matrix of determinant $1$, its characteristic polynomial equals $\Lambda^2-\Lambda \,\textrm{tr}(\mathcal{M}) + 1$. Therefore, for a stable periodic orbit, the stability condition is equivalent to
\begin{equation} -2 < \textrm{tr}(\mathcal{M}) < 2 \, . \end{equation}
For the \emph{axial} periodic orbit of a concave-concave Fabry-Perot cavity, $\textrm{tr}(\mathcal{M}) = 4 g_1 g_2 - 2$, with the usual definition $g_i = 1 - L / R_i$, which leads to the well-known stability condition
\begin{equation} \label{eq:stabilite-axial}
 0 < g_1 g_2 < 1 \, .
\end{equation}
In a given plano-concave Fabry-Perot cavity, there are many different periodic orbits beyond the axial orbit, with different stabilities. The following paragraph gives a few examples.

\subsection{Periodic orbits in plano-concave cavities}\label{sec:orbites-plano-concave}

Some periodic orbits in a plano-concave Fabry-Perot cavity are drawn in Fig.~\ref{fig:differentes-orbites}. For the axial periodic orbit [continuous red line in Fig.~\ref{fig:differentes-orbites}(a)], the stability criterion (\ref{eq:stabilite-axial}) yields the condition
\begin{equation} 0 < \frac{L}{R} < 1 \, . \end{equation}
For $L = R$, this orbit is marginally stable, while it is unstable for $L > R$. In this article, we consider only the case $L < R$ where this orbit is stable.\\

The dotted blue periodic orbit in Fig.~\ref{fig:differentes-orbites}(a), called $C$-orbit, exists for $L / R = \rho > 1 - 1 / \sqrt{2} \simeq 0.29$ and is unstable until $\rho = 1 - \sqrt{2} / 4 \simeq 0.65$. At $\rho \simeq 0.65$, it is marginally stable, and it becomes stable for $\rho > 0.65$. This orbit is not included in the paraxial approximation, but it actually plays a prominent role in the microlaser experiments, see Sec.~\ref{sec:exp-0.55}.\\

The periodic orbit in Fig.~\ref{fig:differentes-orbites}(b), called $W$-orbit, exists only for $\rho = 1 / 2$ and is marginally stable. It exists and has the same length $\mathcal{L}=8L$ for each angle $\alpha$ between $0^{\circ}$ and $60^{\circ}$, i.e., it belongs to a continuous family.\\

Many other periodic orbits of the plano-concave Fabry-Perot cavity are presented in Ref.~\cite{disque-coupe-reichl}.

\subsection{Trace formula for plane concave Fabry-Perot cavities}\label{sec:trace-plan-concave}

In the framework of paraxial approximation, the eigenfrequencies of the plano-concave Fabry-Perot cavity are known explicitly. These can be used to explicitly derive the corresponding trace formula, i.e., express the density of states as a sum over periodic orbits like in Eq.~(\ref{eq:trace-definition}).\\

The eigen-wavenumbers are written [see Ref.~\cite{Siegman1986} Eq.~(19.23), and Part I Eq.~(28)]:
\begin{equation}\label{eq:frequences-propres}
k_{q, p, p'} = \frac{\pi}{L} \, q+(p+p'+1) \frac{\arccos(\sqrt{1-\rho})}{L}
\end{equation}
They depend on three natural numbers, $q$, $p$ and $p'$ where $q$ corresponds to the longitudinal excitation and $p, \, p'$ to the transverse ones. For convenience, we consider $p' = 0$ which is equivalent to a bi-dimensionnal approximation, consistent with the microlaser experiments. The density of states is
\begin{equation}\label{eq:trace-debut}
d(k) = \sum_{p = 0}^\infty \sum_{q = 0}^\infty \delta(k - k_{q,p}) \, .
\end{equation}
if we take an infinite number of transverse excitations into account. The calculation is then not straightforward and the interpretation of the results requires concepts from the theory of dynamical system which are beyond the scope of this paper. Moreover, it is not meaningful to consider transverse excitations of arbitrarily high order since the paraxial approximation breaks down at some point. In real experiments like those presented in Part I and in this article, $q$ is greater than 1000 whereas $p$ is of the order of unity. Hence we consider only a finite number of transverse excitations $p = 0 \ldots p_{max}$ which also greatly simplifies the calculations. We now compute the density of states (\ref{eq:trace-debut}), where we use the Poisson formula (see \ref{sec:annexe-Poisson}) for the summation over $q$, yielding
\begin{eqnarray}
d(k) & = & \sum_{p=0}^{p_{max}} \sum_{Q=-\infty}^\infty \int_0^{\infty}dq \, \delta(k - k_{q,p}) \, e^{2 i \pi Qq} \\
 & = & \frac{L}{\pi} \sum_{p=0}^{p_{max}}\sum_{Q=-\infty}^\infty\, e^{2i Q [Lk - (p+1) \beta]}
\end{eqnarray}
with $\beta=\arccos\sqrt{1-\rho}$. The summation over $p$ is then performed as a geometric series, with $p_m=p_{max}+1$, the total number of $p$ values:
\begin{eqnarray}
d(k) & = & \frac{L}{\pi} \sum_{Q = -\infty}^\infty \, e^{2i Q(Lk - \beta)} \sum_{p=0}^{p_{max}}\, \left( e^{-2i Q\beta} \right)^p \label{eq:serie-divergente} \\
 & = & \frac{L}{\pi} \sum_{Q=-\infty}^\infty \, e^{2i Q(Lk-\beta)}\frac{1-e^{-2i Q\beta p_m}}{1-e^{-2i Q\beta}}\\
  & = & \frac{L}{\pi} \sum_{Q=-\infty}^\infty \, e^{iQ(2Lk-\beta-\beta p_m)}\frac{\sin (Q\beta p_m)}{\sin (Q\beta)}\label{eq:trace-serie-geo}
\end{eqnarray}
The term $Q=0$, so-called \emph{Weyl term}, describes the mean density of states. From Eq.~(\ref{eq:trace-serie-geo}), it is
\begin{equation}
d_{weyl}=\frac{L}{\pi}\,p_m
\end{equation}
We do not consider it anymore to focus on the fluctuating part of the spectrum, i.e. with $Q\neq 0$
\begin{equation}
\tilde{d}(k)=\frac{L}{\pi} \sum_{Q>0}\, 2\cos(2LQk-Q\beta-Q\beta p_m)\,\frac{\sin (Q\beta p_m)}{\sin (Q\beta)}
\end{equation}
Using the elementary formulas $2\sin a\cos b = \sin(a+b)+\sin(a-b)$ and $\sin a = \cos(a-\pi/2)$, the final expression is
\begin{equation} \label{eq:trace-finale}
\tilde{d}(k) = \frac{L}{\pi} \sum_{Q>0}\, \frac{\cos(2LQk-Q\beta-\pi/2)}{\sin (Q\beta)}+ \frac{\cos(2LQk-Q\beta-2Q\beta p_m-\pi/2)}{\sin (Q\beta)}
\end{equation}
which is more or less similar to formula (\ref{eq:trace-definition}). If $\beta=0$, the expression of the trace formula for the plane-parallel Fabry-Perot cavity Eq.~(\ref{eq:trace-plan-plan}) is recovered. The first term in the numerator, $2LQk$, corresponds to the $Q^{th}$ repetition of the axial periodic orbit of length $\mathcal{L}=2LQ$.  The second term in the numerator, $Q\beta=Q\arccos(\sqrt{1-\rho})$, is connected to the Gouy phase.\\

The denominator of expression (\ref{eq:trace-finale}) is equivalent to expression (\ref{eq:cp-stable}). Actually, for a stable periodic orbit, the coefficient of the trace formula is given by $c_{po}\propto 1/\sin(Q\Phi / 2)$, the phase $\Phi$ being defined via the trace of the monodromy matrix:
\begin{equation} \textrm{tr}(\mathcal{M}) = \Lambda_+ + \Lambda_- = e^{i\Phi} + e^{-i\Phi} = 2 \cos\Phi \, . \end{equation}
We have seen that for the axial periodic orbit $\textrm{tr}(\mathcal{M}) = 4 g_1 g_2 - 2 = 2\,(1 - 2\rho)$, which means that
\begin{equation} \label{eq:Phi}
\cos\Phi = 1 - 2\rho \hspace{0.5cm} \Rightarrow \hspace{0.5cm} \Phi = \arccos(1 - 2\rho) = 2 \arccos(\sqrt{1-\rho})
\end{equation}
where the last identity comes from elementary trigonometric calculations. Then the denominator of expression (\ref{eq:trace-finale}) is strictly equivalent to the general expression (\ref{eq:cp-stable}) for a stable periodic orbit.\\

Hence we have demonstrated, that, within the paraxial approximation, the fluctuating part of the density of states can be expressed as a sum over the $Q$ repetitions of the single axial periodic orbit. It should be noted, however, that other periodic orbits of the plano-concave cavity do not appear since our calculation is based on the paraxial approximation. They would be recovered if the full spectrum of the cavitiy, i.e., including modes beyond the paraxial approximation, was considered.

\subsection{The degeneracy condition} \label{sec:trace-degenerescence}

According to expressions (\ref{eq:cp-stable}), the $Q^{th}$ repetition of the axial orbit is weighted by a coefficient $c_{po} \propto 1 / \sin(Q \Phi / 2)$, which diverges if
\begin{equation}
\frac{Q\Phi}{2} =Q\beta= m \pi \hspace{1 cm} m \in \mathbb{N} \, .
\end{equation}
This yields the divergence condition
\begin{equation} \label{eq:rho-divergence}
\rho = \sin^2 \left( \frac{m\,\pi}{Q} \right)
\end{equation}
which is identical to the degeneracy condition given in part I, Eq.~(39). There is actually no divergence. In the trace formula Eq.~(\ref{eq:trace-serie-geo}), if $Q\beta=m\pi$, then the
ratio of the geometric series over $p$ equals 1:
\begin{equation}
\sum_{p=0}^{p_{max}}(e^{-2iQ\beta})^p=\sum_{p=0}^{p_{max}}1=p_m
\end{equation}
The amplitude of the $Q^{th}$ repetition of the axial periodic orbit is then maximum, and not diverging. Moreover, from the analysis of Sec.~\ref{sec:stabilite}, the $Q^{th}$ repetition of the axial periodic orbit is a marginally stable periodic orbit.  In experiments, the $Q^{th}$ harmonic of the Fourier transform is enhanced, as it will be shown in the next section.

\paragraph{Conclusion} The stability criteria were defined from the monodromy matrix, which reduced to the usual ABCD matrix in the paraxial approximation. The trace formula of the plane-concave cavity was then analytically derived from the eigenfrequencies in the paraxial approximation. It involved only the axial periodic orbit and its repetitions. It is enhanced for specific ratios $\rho=L/R$, where there is a degeneracy in accordance with the terminology of Part I. From a geometrical point of view, the axial periodic orbit is stable in the range $0<\rho<1$ and some of its repetitions are marginally stable at these specific ratios.

\section{Experimental results} \label{sec:experiences}

To illustrate the above theoretical issues, this section deals with plano-concave microlasers with several ratios $\rho = L / R$. It must be pointed out that the non-linearities due to lasing are not prominent close to the lasing threshold. Hence the model developed above remains valid even though it is based only on calculations of the passive cavity resonances. The polarization of the pump beam can play a role \cite{PRA-polarisation}, however, for the sake of clarity, this issue is not mentioned hereafter. \\

The first case, $\rho = 0.55$, is generic, while the two other cases, $\rho = 1/2$ and $\rho = 3/4$, correspond to cavity sizes where an enhancement is expected according to formula (\ref{eq:rho-divergence}). From the point of view of Part I, $\rho = 1/2$ and $3 / 4$ present a degeneracy, but not $\rho = 0.55$.

\subsection{Generic case: $L/R=0.55$}\label{sec:exp-0.55}

First, we consider the ratio $\rho = 0.55$, which does \emph{not} fullfill the degeneracy condition (\ref{eq:rho-divergence}). The lasing emission is recorded along the axis, i.e., in the $\theta = 0^{\circ}$ and $180^{\circ}$ directions [see Fig.~\ref{fig:geomAndSetup}(a) for notations].

\begin{figure}
\begin{center}
\includegraphics[width = 1\linewidth]{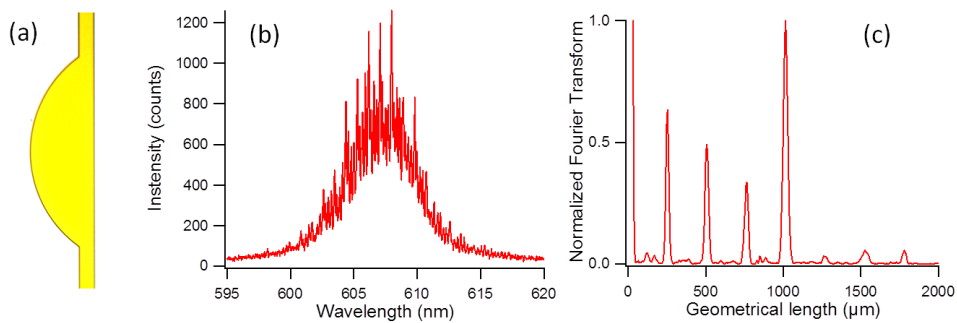}
\end{center}
\caption{Plano-concave Fabry-Perot microlaser with $\rho = 0.55$. (a) Photograph with an optical microscope in real colors. (b) Experimental spectrum for $L = 125~\mu$m in the direction $0^{\circ}$ with a pump polarization orthogonal to the optical axis. The bell-shaped baseline is due to the high density of peaks that is beyond the spectrometer resolution. (c) Fourier Transform of the spectrum in (b).}
\label{fig:photo-0.55}
\end{figure}

Figure \ref{fig:photo-0.55}(b) presents a comb-like spectrum. Its Fourier transform in Fig.~\ref{fig:photo-0.55}(c) evidences a first peak at a length\footnote{In section \ref{sec:experiences}, all the $x$-axis of the Fourier transform are normalized by $1.63$, the global refractive index for these samples. Then, the $x$-axis corresponds to the geometrical length.} $\mathcal{L} = 250~\mu$m, which corresponds to the length of the axial periodic orbit, $2 \cdot L = 250~\mu$m. In Fig.~\ref{fig:photo-0.55}(c), the $4^{th}$ harmonic is greater than the other ones, which is consistent with formula (\ref{eq:cp-stable}). Actually, the following table indicates the relative amplitude of the repetitions of the axial orbit, and the fourth repetition dominates clearly:
\begin{center}
\begin{tabular}{|c|c|c|c|c|c|}
  \hline
  Repetition of the periodic orbit,
$Q$ & 1, fundamental & 2  & 3 & 4 & 5 \\
  \hline
 Relative amplitude from
  formula (\ref{eq:cp-stable}) & 0.67 & 0.50 & 0.84 & 2.21 & 0.58 \\
  \hline
\end{tabular}
\end{center}
From this table, we could expect the $5$th harmonic to be as high as the $2$nd harmonic, but the amplitude of the harmonics decreases in general quite fast with their order, due to the finite linewidth of the resonances [cf.\ the typical example of the plane-parallel Fabry-Perot cavity in Fig.~\ref{fig:plan-plan}(c)].\\

\begin{figure}
\begin{center}
\includegraphics[width=1\linewidth]{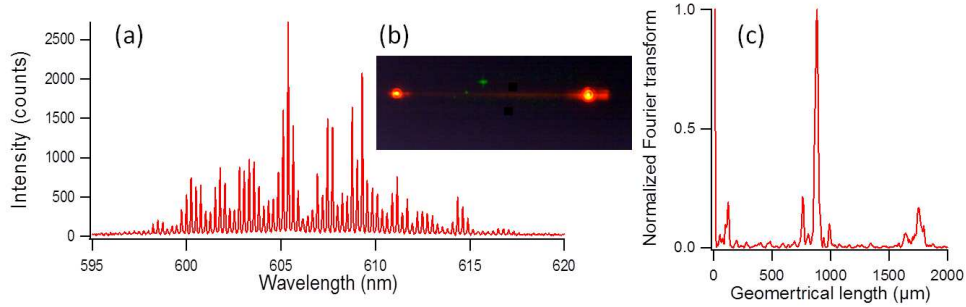}
\end{center}
\caption{Plano-concave Fabry-Perot microlaser with $\rho = 0.55$. (a) Experimental spectrum for $L=125~\mu$m in the direction $\theta = 180^{\circ}$ with a pump polarization parallel to the axis. (b) Photograph of the lasing cavity in the cavity plane and in the direction $180^{\circ}$. The boundary is not visible. The lasing emission is red. The weak green spots stem from the scattering of the pump beam. (c) Fourier transform of the spectrum in (a).}
\label{fig:photo-0.55-orbite-droite}
\end{figure}

When looking into the direction $180^{\circ}$, the spectrum shown in Fig.~\ref{fig:photo-0.55-orbite-droite}(a) is also comb-like, but dramatically different otherwise. Its Fourier transform in Fig.~\ref{fig:photo-0.55-orbite-droite}(c) is peaked at a length of $876~\mu$m, in good agreement with the length of the $C$-orbit shown in Fig.~\ref{fig:differentes-orbites}(c), $\mathcal{L} = 877~\mu$m. This periodic orbit emits only in the direction $180^{\circ}$ since the two reflections on the rounded part of the boundary are confined by total internal reflection. From the monodromy matrix theory presented in section \ref{sec:stabilite}, it follows that this orbit is unstable for $\rho=0.55$. 

Being clearly beyond the paraxial approximation, this orbit does not appear in the model of Section \ref{sec:trace-plan-concave}. But it exists and, according to this experiment, it appears that modes are able to localize on it. The photo of Fig. \ref{fig:photo-0.55-orbite-droite}(b) is registered close to the cavity plane and in the direction $180^{\circ}$. The two lateral spots are precisely localized where they are expected, i.e., at points $A$ and $B$ in Fig.~\ref{fig:differentes-orbites}(a). Following the seminal paper Ref.~\cite{heller1984}, these modes localized on an unstable periodic orbit are called \emph{scars}.\\

To summarize the results for the generic case $\rho=0.55$, two main periodic orbits are identified that sustain lasing modes: the axial orbit the harmonics of which have amplitudes according to formula (\ref{eq:cp-stable}), and the $C$-orbit which is unstable and beyond the paraxial approximation.

\subsection{Case $L/R = 1/2$} \label{sec:exp-0.5}

The ratio $\rho = 1/2$ corresponds to a resonator shape where a degeneracy is expected according to formula (\ref{eq:rho-divergence}). 

\begin{figure}
\begin{center}
\includegraphics[width=1\linewidth]{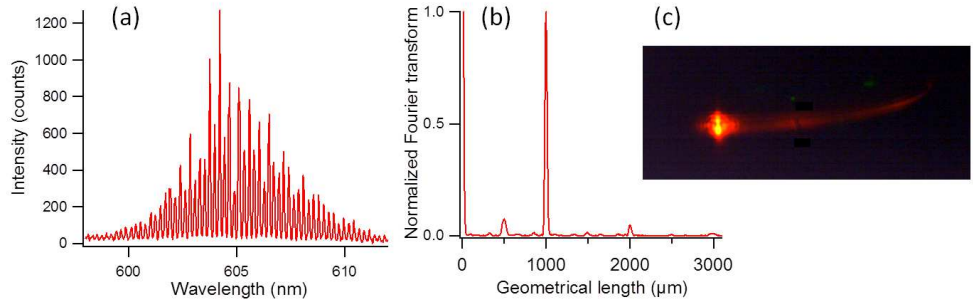}
\end{center}
\caption{Plano-concave Fabry-Perot microlaser with $\rho = 1/2$. (a) Spectrum collected in the direction $\theta = 50^{\circ}$. (b) Fourier transform of the spectrum in (a). (c) Photograph of the lasing cavity from the direction $\theta \simeq 50^{\circ}$. The curved boundary is slightly visible due to the scattering of the red lasing light.}
\label{fig:spectre-0.5}
\end{figure}

The $C$-orbit is still lasing, like in Sec.~\ref{sec:exp-0.55}, but not the axial orbit. For this ratio, Eq.~(\ref{eq:cp-stable}) predicts that the $4^{th}$ repetition of the axial orbit is enhanced. The $W$-orbit of Fig.~\ref{fig:differentes-orbites}(b) is marginally stable and reduces precisely to four times the axial orbit for $\alpha = 0$. The refractive losses of the $W$-orbit are smaller than those of the axial orbit if $\alpha$ is large enough such that the reflections on the plane cavity boundary [points $A$ and $B$ in Fig.~\ref{fig:differentes-orbites}(b)] are totally reflected. Two bright spots are indeed observed in the directions $\theta \simeq \pm \, 50^{\circ}$ as shown in Fig.~\ref{fig:spectre-0.5}(c). The corresponding spectra are comb-like and their Fourier transform is peaked at $\mathcal{L} = 999~\mu$m, as presented in Fig.~\ref{fig:spectre-0.5}(b), which is fully consistent with the length of the $W$-orbit, $\mathcal{L} = 4 \cdot 2 \cdot L = 1 000~\mu$m. 

\subsection{Another degenerate case: $L/R = 3/4$} \label{sec:exp-0.75}

The observations for $\rho = 1/2$ are in agreement with the theory presented in this paper, but they are not generic since the $W$-orbit exists only at this particular ratio. This section deals with the case of $L/R = 3/4$, which corresponds to a degenerate cavity as discussed in Part I, and is generic. 

\begin{figure}
\begin{center}
\includegraphics[width=1\linewidth]{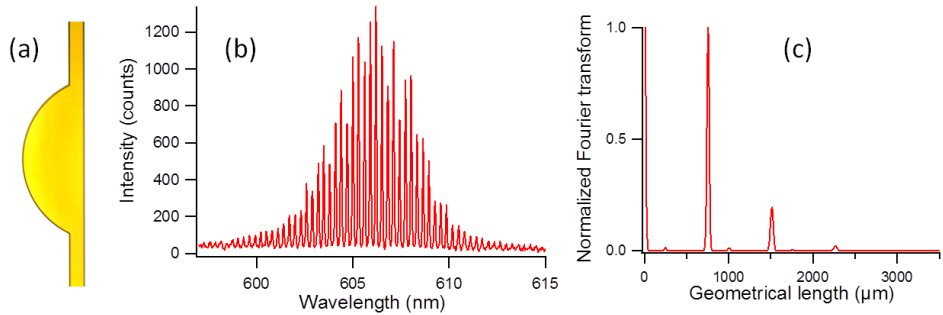}
\end{center}
\caption{Plano-concave Fabry-Perot microlaser with $\rho = 3/4$. (a) Photograph with an optical microscope in real colors. (b) Experimental spectrum for $L=125~\mu$m in the direction $\theta = 0^{\circ}$. (c) Fourier transform of the spectrum in (b).}
\label{fig:spectre-0.75}
\end{figure}

According to formula (\ref{eq:cp-stable}), the third harmonic of the axial orbit is expected to be enhanced for this case. In Fig.~\ref{fig:spectre-0.75}(c), the Fourier transform of the spectrum along the direction $\theta = 0^{\circ}$ is peaked at $755~\mu$m, which agrees very well with the theoretical length of $\mathcal{L} = 3 \cdot 2 \cdot 125 = 750~\mu$m.\\

The three examples considered in the Section ($\rho=\frac{1}{2}, \frac{3}{4}$, and 0.55) demonstrate a good agreement between experiments and the theory based on the trace formula. Each prediction which can reasonably be checked is indeed confirmed experimentally.

\section{Conclusion}

This paper illustrates the potential of semi-classical physics with the example of the well-known Fabry-Perot cavity. The periodic orbits are the basis of any semi-classical description. Here, the derivation of the trace formula demonstrates that only the axial periodic orbit and its repetitions are involved in the paraxial approximation. On the contrary, experiments evidence that lasing modes can be localized on the $C$-orbit, which is unstable and not included in the paraxial approximation.\\

The semi-classical physics aims at highlighting the correspondence between the wave and the geometrical points of view. Here, the trace formula makes the connection between the mode degeneracy discussed in Part I, and the change of stability of the repetitions of the axial periodic orbit. These techniques are but a part of the powerful toolbox of semi-classical physics, which is relevant in various domains, especially for laser cavities.\\

\section*{Acknowledgments}

The authors, especially M. L., are earnestly grateful to the PhD thesis of N. Barr\'e. In particular, Fig.~10 in Part I was highly inspiring. The authors acknowledge the Renatech facilities. M. L. would like to dedicate this paper to J.-F. Roch.\\

\appendix

\section{Poisson summation formula}\label{sec:annexe-Poisson}

The usual Poisson summation formula is
\begin{equation}
\sum_{q=-\infty}^\infty f(k-k_q) = \sum_{Q=-\infty}^\infty \int_{-\infty}^{\infty} dq \, f(k-k_q) \, e^{2 \pi i Qq}
\end{equation}
where $f$ is a function or a distribution and the sum in the left hand term starts from $q=-\infty$. However, in the case we are considering, the longitudinal number $q$ is only positive. The Poisson formula is then accordingly modified to
\begin{equation}\label{eq:Poisson}
\sum_{q=0}^\infty f(k-k_q) = \sum_{Q=-\infty}^\infty \int_0^{\infty} dq \, f(k-k_q) \, e^{2 \pi i Qq} + \frac{1}{2} f(k-k_0)
\end{equation}
following Eq.~(2.157) in Ref.~\cite{brack}. The second part of the right hand term corresponds to a longitudinal number $q = 0$, and hence to minor corrections which are discarded here.

\end{document}